\documentclass[reprint,aps,pra,longbibliography, groupedaddress,nofootinbib]{revtex4-1}
\usepackage{graphicx,amsmath,amssymb,braket,siunitx,mathtools,mathrsfs}
\usepackage[dvipsnames]{xcolor}
\usepackage[unicode]{hyperref}
\usepackage{cleveref}
\usepackage{oplotsymbl}
\DeclareSIUnit\gauss{G}

\DeclarePairedDelimiterXPP\BigOSI[2]%
  {\mathcal{O}}{(}{)}{}%
  {\SI{#1}{#2}}

\hypersetup{
    colorlinks=true,
    citecolor=Blue,
    linkcolor=Blue,
    urlcolor=Blue
}
\newcommand{\mref}[2]{%
    \hyperref[{#1}]{%
        \ref*{#1}#2%
    }%
}
\begin{document}
\title{A general approach to state-dependent optical tweezer traps for polar molecules}
\author{L. Caldwell}
\email[]{l.caldwell@imperial.ac.uk}
\altaffiliation[Present Address: ]
{JILA, NIST and University of Colorado,
Boulder, Colorado 80309-0440, USA.}
\author{M. R. Tarbutt}
\email[]{m.tarbutt@imperial.ac.uk}
\affiliation{Centre for Cold Matter, Blackett Laboratory, Imperial College London, Prince Consort Road, London SW7 2AZ UK
}

\begin{abstract}
    State-dependent optical tweezers can be used to trap a pair of molecules with a separation much smaller than the wavelength of the trapping light, greatly enhancing the dipole-dipole interaction between them. 
    Here we describe a general approach to producing these state-dependent potentials using the tensor part of the ac Stark shift and show how it can be used to carry out two-qubit gates between pairs of molecules.
    The method is applicable to broad classes of molecules including bialkali molecules produced by atom association and those amenable to direct laser cooling.
\end{abstract}

\maketitle
\begin{figure*}
    \centering
    \includegraphics{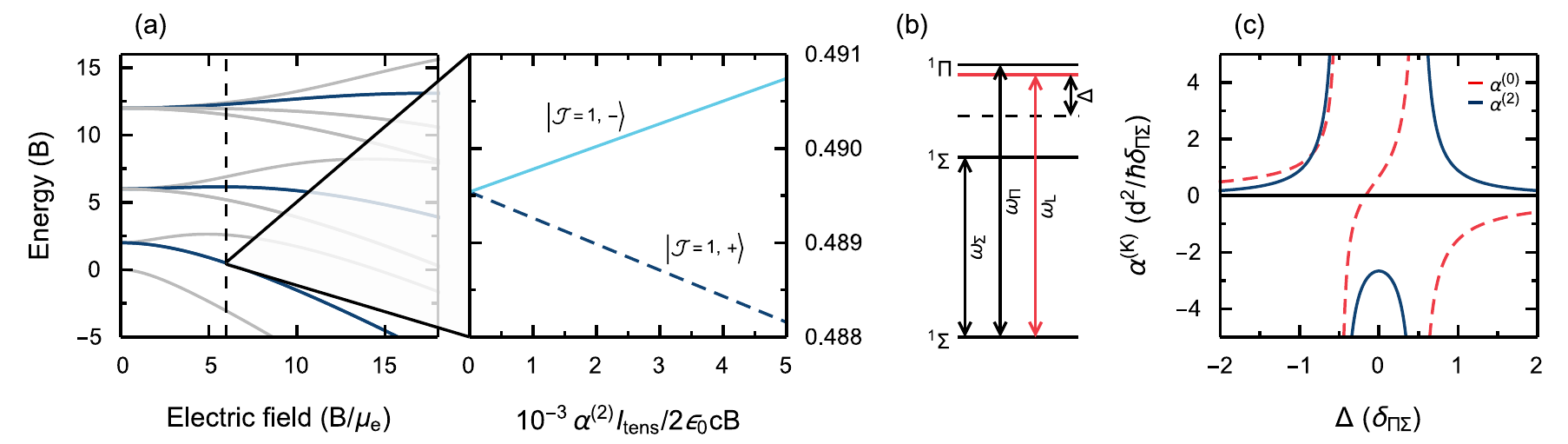}
    \caption{Level shifts of a molecule with a $^1\Sigma$ ground state in dc and laser fields. (a) Energy of states as a function of electric field. States with $|m_J|=1$ are shown in blue. Inset: states in the $\ket{\mathcal{J}=1, m_J=\pm1}$ manifold as a function of tensor tweezer intensity, $I_{\rm tens}$. Calculated for $E_z = 6B/\mu_e$ with $\hat{\epsilon}=\hat{x}$ and $\alpha^{(2)} <0$. (b) Schematic of relevant electronic structure in the molecule. (c) Scalar (red dashed) and tensor polarizabilities (blue) calculated using Eqs.~(\ref{eq:alpha0}) and (\ref{eq:alpha2}) with $d_{\Sigma}=d_{\Pi}=d$.}
    \label{fig:structure}
\end{figure*}

\section{Introduction}

An array of polar molecules formed using optical tweezer traps is an attractive platform for quantum simulation, quantum information processing, and the study and control of collisions and chemical reactions at ultracold temperatures. Long range electric dipole-dipole interactions between the molecules can be used to implement quantum gates~\cite{DeMille2002, Yelin2006, Pellegrini2011, Ni2018, Hudson2018, Hughes2020, Caldwell2020c} and a wide variety of Hamiltonians that model interesting many-body quantum systems~\cite{Micheli2006, Gorshkov2011, Wall2010, Wall2013, Yan2013, Hazzard2013, Hazzard2014}. Tweezer arrays are reconfigurable~\cite{Endres2016, Barredo2018}, so arbitrary geometries can be formed, defects can be removed or deliberately implanted, and interactions targeted to selected molecules by bringing them close to each other. There has been rapid recent progress in realizing such molecular arrays. Single NaCs molecules have been formed by association of Na and Cs atoms inside a tweezer trap~\cite{Liu2018,Liu2019, Zhang2020}, and single molecules of $^{85}$Rb$^{87}$Rb have been formed in a tweezer by microwave association assisted by coupling of the spin and motional degrees of freedom~\cite{He2020}. A tweezer array of laser-cooled CaF has been demonstrated and used to study collisions between pairs of molecules in selected states~\cite{Anderegg2019, Cheuk2020}. 

The dipole-dipole interaction energy $E_{\rm dd}$, between a pair of molecules in such an array is limited by the minimum separation between adjacent traps. In a conventional tweezer array this is set by the diffraction-limited spot size of each tweezer; roughly the wavelength of the trapping light, typically $\sim\SI{1}{\micro\meter}$, giving $E_{\rm dd}\sim\SI{1}{\kilo\hertz}$. Recently, we introduced the idea of a state-dependent tweezer trap which confines two molecules in the same tweezer with controllable spacing on a sub-wavelength scale~\cite{Caldwell2020c}. The method applies to molecules with electron spin, such as those produced by direct laser cooling, and relies on the interaction between the vector polarizability of the molecule and the polarization gradients around the focus of the tweezer. The tight focussing produces an elliptically polarized field with opposite handedness on either side of the focus. The resulting vector Stark shift depends on this handedness and on the orientation of the molecule’s spin, so two molecules with opposite spin projections are trapped at different locations. Their separation can be controlled by superimposing another tweezer at a different wavelength. This method can enhance the dipole-dipole interaction by two orders of magnitude, enabling much faster and more robust gate operations. The scheme is ultimately limited by photon scattering which is higher than in a standard tweezer because one wavelength has to be tuned between the fine-structure components of an excited state,  resulting in a comparatively small detuning, especially for light molecules. Furthermore, the scheme fails for spinless molecules, which is a major shortcoming since all molecules associated from ultracold alkali atoms are spinless in their ground states.

Here, we present a state-dependent tweezer based on the tensor Stark shift, which is large for most molecules. We produce two separated traps for molecules in two different states by using tweezers at two wavelengths, one where the scalar shift dominates and the other where the tensor shift is dominant. The sub-wavelength separation of the molecules is controlled through the separation and relative intensities of the tweezers. The scheme works for molecules with or without electron spin, greatly reduces the photon scattering rate, and enhances the dipole-dipole interaction by a factor 100 or more. We show how this can be used to implement fast two-qubit gates using electric-field-induced dipoles or resonant dipole-dipole interactions, and discuss the implementation of these ideas using NaCs and CaF molecules. We note that other methods for producing sub-wavelength potentials for molecules have recently been studied~\cite{Kruckenhauser2020}.

\section{Method}

To illustrate our scheme, we consider a simple $^1 \Sigma$ molecule with dipole moment $\mu_e$, rotational constant $B$ and no hyperfine structure. Figure~\mref{fig:structure}{(a)} shows the Stark shift of the electronic ground state in an applied electric field along $z$, $E_z$. The electric field mixes states with different rotational quantum number $J$ but equal projection onto the $z$ axis $m_J$. We label the resulting states $\ket{\mathcal{J},m_J}$ where $\mathcal{J}$ is the rotational level the state connects to at zero field. In the absence of other fields, the states $\ket{\mathcal{J},m_J}$ and $\ket{\mathcal{J},-m_J}$ are degenerate and we refer to these pairs as an $|m_J|$ manifold. Consider the response of this molecule to a light field with polarization $\hat{\epsilon}$ and intensity $I$. We assume that the dc Stark shift is much larger than the ac Stark shift. The molecule-light interaction has scalar, vector and tensor parts whose dependence on the frequency of the light can be factored out into three constants $\alpha^{(0)}$, $\alpha^{(1)}$ and $\alpha^{(2)}$; the scalar, vector and tensor polarizabilities. The scalar part of the interaction shifts every state by $W_0=-\frac{1}{2\epsilon_0 c}\alpha^{(0)} I$. In the absence of any electronic or nuclear spin, $\alpha^{(1)}=0$ so there is no vector part. The tensor part causes the state-dependent shifts which we use in our scheme. 

The values of $\alpha^{(0)}$ and $\alpha^{(2)}$ depend on the electronic structure of the molecule. Here, for simplicity, we assume they are dominated by a single rovibrational level from each of the first two electronic excited states -- a $\Sigma$ state and a $\Pi$ state with energies $\hbar \omega_\Sigma$ and $\hbar \omega_\Pi$ above the ground state, as shown in Fig~\mref{fig:structure}{(b)}. If we define $\Delta$, the detuning of the light from the midpoint of the two states, the polarizabilities can be written\footnote{See for example equations A16 and A18 of Ref.~\cite{Caldwell2020b}.}
\begin{align}
    \alpha^{(0)}\simeq -\frac{1}{3\hbar}\left(\frac{ d_{\Sigma}^2}{\Delta+\frac{\delta_{\Pi\Sigma}}{2}}+\frac{2 d_{\Pi}^2}{\Delta-\frac{\delta_{\Pi\Sigma}}{2}}\right),\label{eq:alpha0}\\
    \alpha^{(2)}\simeq -\frac{2}{3\hbar}\left(\frac{ d_{\Sigma}^2}{\Delta+\frac{\delta_{\Pi\Sigma}}{2}}-\frac{d_{\Pi}^2}{\Delta-\frac{\delta_{\Pi\Sigma}}{2}}\right).\label{eq:alpha2}
\end{align}
Here, $d_\Sigma$ and $d_\Pi$ are the transition dipole moments between the ground state and excited states, $\delta_{\Pi\Sigma}=\omega_\Pi-\omega_\Sigma$ and we have assumed $|\Delta|, |\delta_{\Pi\Sigma}| \ll \omega_\Pi, \omega_\Sigma$. Figure~\mref{fig:structure}{(c)} shows $\alpha^{(0)}$ and $\alpha^{(2)}$ as a function of $\Delta$ in this limit.

Our scheme uses optical tweezers at two different wavelengths. The first, which we call the scalar tweezer, has intensity $I_{\rm sc}$ and wavelength $\lambda_{\rm sc}$ such that $\Delta = \Delta_{\rm sc}$ is negative with $|\Delta_{\rm sc}| > 2|\delta_{\Pi\Sigma}|$. Under these conditions, $|\alpha^{(0)}| \gg |\alpha^{(2)}|$, so the scalar part of the interaction dominates and the shifts are nearly independent of $\hat{\epsilon}$. The second tweezer, which we call the tensor tweezer, has intensity $I_{\rm tens}$ and wavelength $\lambda_{\rm tens}$ such that 
\begin{equation}
    \Delta=\Delta_{\rm tens}= \frac{\delta_{\Pi\Sigma}}{2}\frac{d_\Sigma^2-2d_\Pi^2}{d_\Sigma^2+2d_\Pi^2}.
\end{equation}
Here $\alpha^{(0)}=0$ while $\alpha^{(2)}=-(d_\Sigma^2+2 d_\Pi^2)/\hbar\delta_{\Pi\Sigma}$, whose magnitude can be large and whose sign is negative (positive) when the $\Pi$ state lies above (below) the $\Sigma$ state. The resulting state-dependent shifts depend on the polarization of the tensor tweezer. We choose this to be linearly polarized perpendicular to the electric field, along either $x$ or $y$. The diagonal matrix elements of the molecule-light interaction are equal for the two states of an $|m_J|$ manifold. The off-diagonal part can couple pairs of states with $m_J - m_J'=2$. Due to the electric field, the different $|m_J|$ manifolds are well separated, and the only states coupled by the light are pairs with $|m_J|=1$. These pairs are maximally mixed to give the two states
$
    \ket{\mathcal{J},\pm}=\frac{1}{\sqrt{2}}\left(\ket{\mathcal{J},1}\pm\ket{\mathcal{J},-1}\right)
$
with energies $E_{\mathcal{J},+}$ and $E_{\mathcal{J},-}$. For $\hat{\epsilon}$ along $x$ (along $y$) and $\alpha^{(2)}<0$, $E_{\mathcal{J},-}$ lies above (below). The energy difference $E_{\mathcal{J},+}-E_{\mathcal{J},-}$ is proportional to $I_{\rm tens}$, as shown in the inset of Fig.~\ref{fig:structure}{(a)}, and is similar for all values of $\mathcal{J}$. The off-diagonal matrix element is larger than the diagonal one so that, for each polarization, one state is trapped in the tensor tweezer while the other is repelled. A more complete treatment of the excited states changes the polarizability quantitatively but does not change the key principles essential to our scheme. Importantly, because the tensor tweezer is tuned between different electronic levels rather than between fine-structure components of the same state as in Ref.~\cite{Caldwell2020c}, the photon scattering rate $R_{\rm ph}$ can be much lower for a given trap depth.

Two molecules can be brought to sub-wavelength separations using combinations of scalar and tensor traps. This reduced separation increases the dipole-dipole interaction between them, allowing faster two-qubit operations. We consider two schemes for controlling the separation $r$ of a pair of molecules, one prepared in $\ket{\mathcal{J},+}$ and the other in $\ket{\mathcal{J}',-}$, and show how to implement two-qubit gates in both cases. In one scheme the two trap depths are different, while in the other they are the same. In both cases the relative intensities of the different traps controls the separation while the overall intensity controls the trap frequencies. Throughout, we assume the molecules are cooled to the motional ground states of their traps~\cite{Caldwell2020b}.

\begin{figure}
    \centering
    \includegraphics{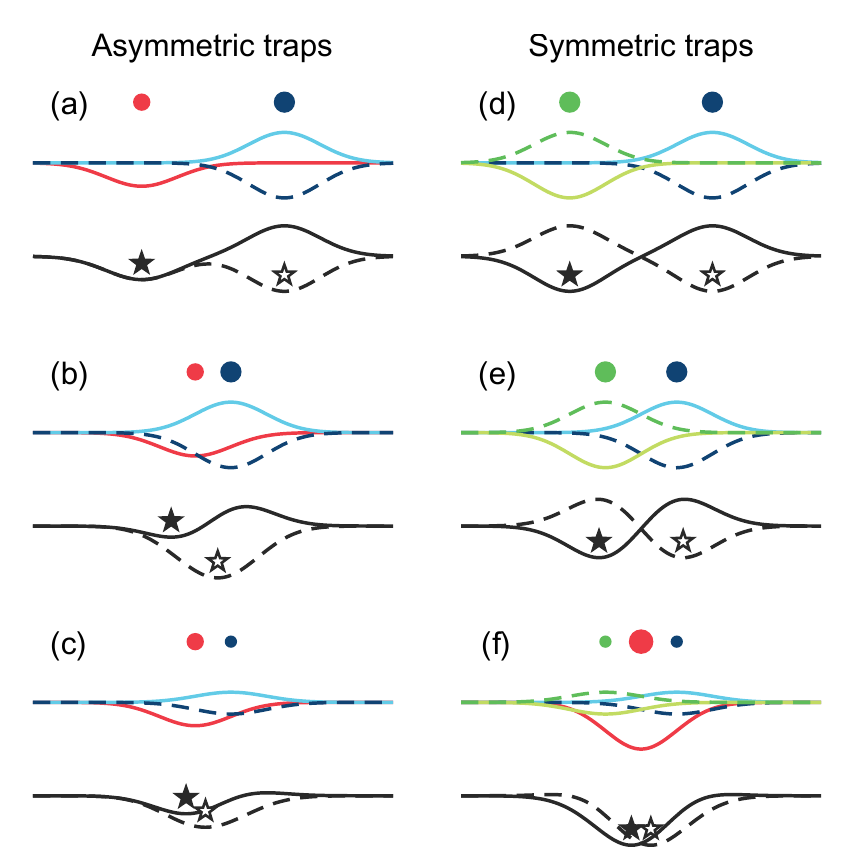}
    \caption{Schematic of the two schemes. Left column (a-c): asymmetric scheme. Right column (d-f): symmetric scheme. (a,d) Separated tweezers. (b,e) Merged tweezers. (c,f) Sub-wavelength spaced tweezers. Dots show positions of tweezers, their size shows intensity and their color shows type -- red for scalar tweezer, blue (green) for tensor tweezer polarized along $x$ ($y$). Solid (dashed) lines show potentials for $\ket{\mathcal{J},-}$ ($\ket{\mathcal{J}',+}$) states. Black lines (offset for clarity) show the combined potential of all tweezers and the filled (open) stars show the positions of the $\ket{\mathcal{J},-}$ ($\ket{\mathcal{J}',+}$) states at each step.}
    \label{fig:basic-schemes}
\end{figure}

\begin{figure}
    \centering
    \includegraphics{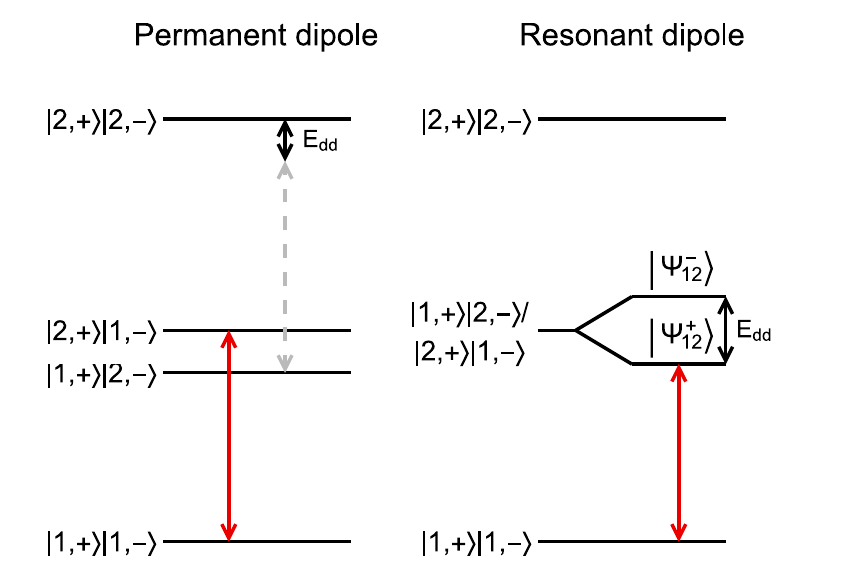}
    \caption{Schematic of the two qubit gate operations. Left side shows gate using permanent dipole moment and right side shows entanglement operation using resonant dipole-dipole interaction.}
    \label{fig:gates}
\end{figure}

\subsection{Asymmetric traps}

The first scheme is illustrated in the left hand column of Fig.~\ref{fig:basic-schemes}. The $\ket{\mathcal{J},+}$ molecule is loaded into a scalar trap and the $\ket{\mathcal{J}',-}$ molecule into a tensor trap with $\hat{\epsilon}$ along $x$. The two tweezers are brought together until the maximum intensity gradient from the tensor tweezer coincides with the center of the scalar tweezer. This can be done without tunneling between sites by ensuring the tensor tweezer is significantly deeper than the scalar tweezer. Once they reach this position, the separation between the potential minima for the two states is controlled by varying the relative intensity of the two tweezers. 

This scheme is suitable for implementing a gate that uses electric dipole moments induced by a static electric field~\cite{DeMille2002}. An example is shown on the left-hand side of Fig.~\ref{fig:gates}. We take $\ket{1,\pm}$ and $\ket{2,\pm}$ as our computational basis states. Their dipole moments, $\mu_1$ and $\mu_2$, can be very different at electric fields that are quite easy to reach for most polar molecules. For example, when $E_z=6B/\mu_e$ (dashed line in Fig.~\mref{fig:structure}{(a)}), $\delta\mu=\mu_1-\mu_2=0.43 \mu_e$. Single qubit operations are driven by microwaves polarized along $z$. Because the two trap depths are different, and the frequency of the transition $\ket{1,+}\leftrightarrow\ket{2,+}$ has a different gradient with intensity from that of $\ket{1,-}\leftrightarrow\ket{2,-}$, the two qubits can be addressed separately. A CNOT gate can be implemented by driving a microwave transition resonant with $\ket{1,+}\ket{1,-}\leftrightarrow\ket{2,+}\ket{1,-}$ with sufficient duration to resolve it from $\ket{1,+}\ket{2,-}\leftrightarrow\ket{2,+}\ket{2,-}$. Their frequencies differ by $E_{\rm dd}/h=\delta\mu^2/(4 \pi\epsilon_0 h |r^3|)$ due to the dipole-dipole interaction.

\subsection{Symmetric traps} 

The second scheme for reaching sub-wavelength separation is illustrated in the right hand column of Fig.~\ref{fig:basic-schemes}. The $\ket{\mathcal{J},-}$ molecule is loaded into a tensor tweezer with polarization along $x$ and the $\ket{\mathcal{J}',+}$ molecule into another tensor tweezer with polarization along $y$. The two tweezers are brought together until they are separated by $\sim \lambda_{\rm tens}$ such that the maximum gradients of the two potentials coincide\footnote{Note that the two tweezers should have optical frequencies that are different by at least a few \si{\mega\hertz} to avoid interference effects between the overlapping  beams. This small difference will have a negligible effect on the polarizabilities.}. We then turn on a scalar tweezer located at the midpoint of the two tensor tweezers which squeezes the two molecules together. The separation can be tuned by varying the relative intensity of the scalar and tensor tweezers. 

Two-qubit gates using permanent dipole moments can be carried out in the same way as described above. In addition, due to the symmetry of this scheme, the resonant dipole-dipole interaction can be utilized as shown on the right-hand side of Fig.~\ref{fig:gates}. As before, let us take $\ket{1,\pm}$ and $\ket{2,\pm}$ as our computational basis. If the scalar tweezer has no tensor contribution, the transitions $\ket{1,-}\leftrightarrow \ket{2,-}$ and $\ket{1,+}\leftrightarrow \ket{2,+}$ are exactly degenerate when the intensities of the two tensor tweezers are equal. For realistic parameters the scalar tweezer has a small tensor component, but the degeneracy can be recovered by tuning the relative intensity of the two tensor tweezers to cancel its contribution. The two-qubit states $\ket{1,-}\ket{2,+}$ and $\ket{2,-}\ket{1,+}$ are mixed by the dipole-dipole interaction, and the new eigenstates are the entangled states $\ket{\Psi_{12}^{\pm}}=\tfrac{1}{\sqrt{2}}(\ket{1,+}\ket{2,-}\pm\ket{1,-}\ket{2,+}$, separated by the dipole-dipole interaction energy. Two-qubit gates can be carried out by driving transitions which resolve this energy splitting. In this case, the electric field need only be large enough to dominate over the interaction with the light fields. In practice, it can be useful to use a third manifold, say $\ket{3,\pm}$. Then, $\ket{1,\pm}$ and $\ket{3,\pm}$ form the computational basis and for small $E_z$ have negligible transition dipole moment between them, so are (very nearly) stationary states, while $\ket{2,\pm}$ is used to implement the gate as described above. Suitable gate protocols are described in detail in Refs.~\cite{Ni2018, Caldwell2020c}. To carry out single qubit operations, the intensities of the two tensor tweezers can be unbalanced so that the individual qubit frequencies can be resolved.

\section{Minimum separation} 

The schemes illustrated in Fig.~\ref{fig:basic-schemes} can bring the two traps arbitrarily close together, but eventually the wavefunctions of the two molecules overlap and there are inelastic collisions. We consider a pair of molecules in the motional ground state of tweezers with radial and axial trap frequencies $\omega_{r}$ and $\omega_z$ and associated harmonic oscillator lengths of $r_0$ and $z_0$. For unpolarized molecules the long range interaction, $-C_6/r^6$, has a negligible influence on the motional wavefunctions as long as $r$ is a few times greater than a critical distance $r_{\rm c} =(2C_6/\hbar \omega_{r})^{1/6}$. For most cases of interest $r_{\rm c} < \SI{50}{\nano\meter}$. In this case, the collision rate is
\begin{equation}
\begin{split}
    R_{\rm col} &= k_2\int |\psi_{\tiny\starletfill}(x)|^2|\psi_{\tiny\starlet}(x)|^2 \mathrm{d}^3 x,\\ &= \frac{k_2}{(2\pi)^{3/2}r_0^2 z_0}e^{-r^2/(2r_0^2)},
    \label{eq:Gamma_coll}
\end{split}
\end{equation}
where $\psi_{\tiny\starletfill}$, $\psi_{\tiny\starlet}$ are the wavefunctions of the two molecules and $k_2$ is the inelastic rate coefficient. Although $k_2$ can vary greatly from one system to another, several experiments (e.g. \cite{Ospelkaus2010, Gregory2019, Yang2019, Cheuk2020}) have measured values close to that predicted by a simple single channel model with unity probability of loss at short range. For distinguishable particles near zero energy, this universal loss coefficient is~\cite{Idziaszek2010}
\begin{equation}
    k_2 = \frac{8 \pi^2 }{\Gamma(1/4)^2}\left( \frac{2\hbar^2C_6}{m_{\rm red}^3}\right)^{1/4},
    \label{eq:beta}
\end{equation}
where $m_{\rm red}$ is the reduced mass. For molecules with substantial dipole moments, $C_6$ is dominated by a rotational contribution giving $C_6 \approx 1/(4\pi\epsilon_0)^2 \times \mu_{\rm e}^4/(6B)$~\cite{Julienne2011,Zuchowski2013}. Taking characteristic values of $\mu_{\rm e} = 3$~D, $B/h = \SI{5}{\giga\hertz}$, $m_{\rm red} = 50$~amu, $\omega_r/2\pi = 100$~kHz and $\omega_z/2\pi = 23$~kHz, Eqns.~(\ref{eq:Gamma_coll}) and (\ref{eq:beta}) give $R_{\rm col}=4$~s$^{-1}$ at $r=200$~nm, rising to 270~s$^{-1}$ at 150~nm. For separations in this range, the dipole-dipole interaction is enhanced by a factor of roughly 100 relative to normal tweezers. An electric field that induces aligned dipoles sets up a repulsive barrier that can greatly suppress $k_2$~\cite{Miranda2011, Julienne2011, Valtolina2020}, so that smaller values of $r$ can be used to generate even larger enhancements of the dipole-dipole interaction.

\section{Implementation with real molecules} 

The nuclear and electronic spins in real molecules couple to the rotational motion, complicating the structure. However, in a modest electric field the Stark interaction dominates over spin-rotation interactions so the rotational motion decouples from the spins. In this limit, we label the states $\ket{\mathcal{J},m_J,G,m_G}$ where $G$ and $m_G$ are the quantum numbers for the total spin and its projection onto the $z$ axis. In the absence of light, $\ket{\mathcal{J},m_J,G,m_G}$ and $\ket{\mathcal{J},-m_J,G,-m_G}$ are degenerate. For $|m_J|=1$ manifolds, the tensor Stark shift mixes $\ket{\mathcal{J},m_J,G,m_G}$ and $\ket{\mathcal{J},-m_J,G,m_G}$ which are not degenerate in general. To obtain the linear shifts used in our scheme we either need the tensor light shifts to dominate over spin-rotation couplings, or need to take $m_G=0$. Below we discuss an example of each case. We note that although hyperfine structure adds complexity, it can also be useful. For example, the two qubits can use different spin manifolds so that the qubit frequencies differ due to the $\mathcal{J}$-dependence of the hyperfine splittings. This does not change the dipole-dipole interaction used for the gates, but is helpful for single-qubit addressability. 

\begin{figure}
    \centering
    \includegraphics{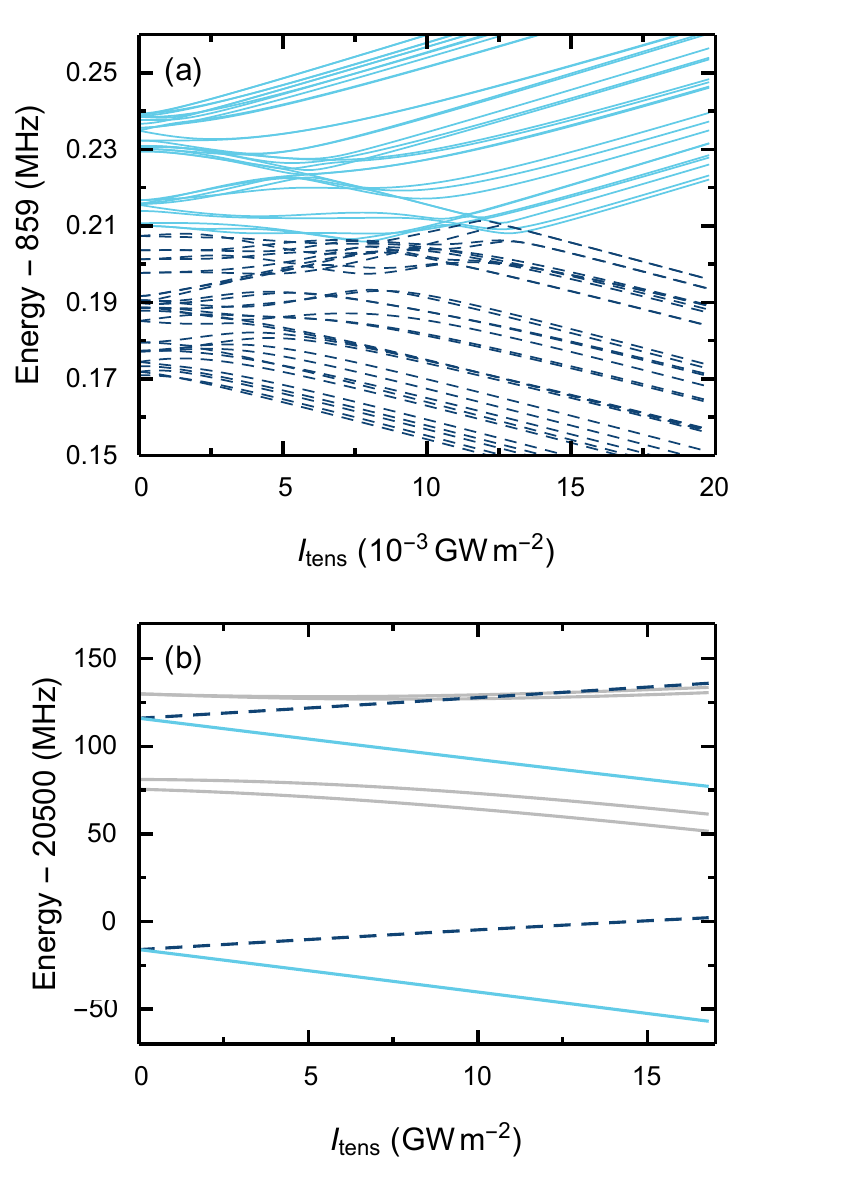}
    \caption{The $\ket{\mathcal{J}=1, m_J=\pm1}$ manifold in real molecules. (a) $^{23}$Na$^{133}$Cs at $E_z = \SI{4.5}{\kilo\volt\per\centi\meter}$, assuming $\tilde{\alpha}^{(2)}=\SI{-8.4e-3}{\hertz\per\watt\meter\squared}$. For intensities above $\sim \SI{0.015}{\giga\watt\per\meter\squared}$ the levels behave as in the model molecule. (b) $^{40}$Ca$^{19}$F at $E_z=\SI{2}{\kilo\volt\per\centi\meter}$ and $\tilde{\alpha}^{(2)}=\SI{6.3e-3}{\hertz\per\watt\meter\squared}$. There are 8 states but the pair with $|m_F|=2$ are nearly degenerate to high intensity. The two pairs with $m_G=0$ split linearly with $I_{\rm tens}$ out to intensities of about \SI{20}{\giga\watt\per\meter\squared}, corresponding to energy splittings of \SI{50}{\mega\hertz}. Light blue (dark blue dashed) lines show states which correspond to $\ket{1,-}$ ($\ket{1,+}$) for the intensities relevant to our scheme. Grey lines are other states in the manifold not useful for our scheme.
    \label{fig:realMolecules}}
\end{figure}

\begin{figure}
    \centering
    \includegraphics{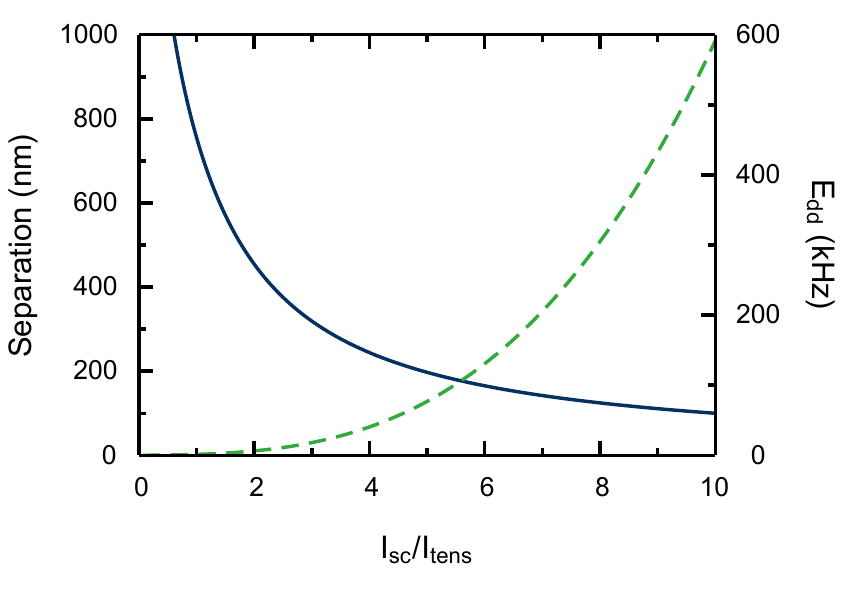}
    \caption{Separation and interaction energy for NaCs. Blue line, left axis: separation of trap minima as a function of $I_{\rm sc}/I_{\rm tens}$. Dashed green line, right axis: dipole-dipole interaction energy for point molecules positioned at trap minima. Calculations are for $\lambda_{\rm sc}=\SI{1064}{\nano\meter}$, $\lambda_{\rm tens}=\SI{768}{\nano\meter}$ with polarizabilities calculated for NaCs as described in the Appendix, and $E_z = \SI{4.5}{\kilo\volt\per\centi\meter}$ giving $\delta\mu=0.43\times4.6\,{\rm D}$.\label{fig:separation}}
\end{figure}

Let us now consider how to implement the asymmetric trap scheme using $^{23}$Na$^{133}$Cs molecules, which have recently been prepared in tweezer traps~\cite{Liu2018,Liu2019, Zhang2020}. The tensor Stark shifts of the $\ket{\mathcal{J}=1, m_J=\pm1}$ manifold are shown in Fig.~\mref{fig:realMolecules}{(a)}. The manifold is split into 64 states due to the interaction of the nuclear spins ($I_{\rm Na}=\tfrac{3}{2}$, $I_{\rm Cs} = \tfrac{7}{2}$) with each other and with the rotational motion. The splittings are small, typically $0.1 - \SI{10}{\kilo\hertz}$, and $I_{\rm tens}$ is large enough to separate the levels out into two groups of 32 states which behave identically to the $\ket{\mathcal{J},\pm}$ states described above.  Each has at least one strong dipole-allowed transition to a state in the $\ket{\mathcal{J}+1,\pm}$ manifold. Many rovibrational levels of each excited electronic state contribute to the polarizability. To avoid high $R_{\rm ph}$, the light must be tuned to a place where the transition strengths to nearby excited states is negligible. In the Appendix, we calculate that a suitable detuning exists at around \SI{768}{\nano\meter}, between the $A^1\Sigma^+$ and $B^1\Pi$ states. We estimate $\tilde{\alpha}^{(2)}=\alpha^{(2)}/2\epsilon_0 h c\simeq\SI{-8.4e-3}{\hertz\per\watt\meter\squared}$, $\alpha^{(0)}\simeq0$, and $R_{\rm ph} \simeq \SI{0.1}{\per\second}$ per \si{\micro\kelvin} of trap depth. Figure~\ref{fig:separation} shows the separation of the trap minima as a function of the ratio $I_{\rm sc}/I_{\rm tens}$. Also shown is the dipole-dipole energy for two fixed point dipoles situated at the trap minima. At small separations, this energy can exceed the energy spacing between motional states in the trap. In this case, it becomes necessary to consider the motional degree of freedom of the molecules when calculating the dipole-dipole interaction energy. A calculation of this kind can be found in Ref.~\cite{Caldwell2020c}. 

Next we consider the implementation of the symmetric trap scheme using $^{40}$Ca$^{19}$F, which has also recently been trapped in tweezers~\cite{Anderegg2019, Cheuk2020}. Figure \mref{fig:realMolecules}{(b)} shows the tensor Stark shifts of the $\ket{\mathcal{J}=1, m_J=\pm1}$ manifold. The interaction of the electronic spin $S=\tfrac{1}{2}$ and nuclear spin $I_{\rm F}=\tfrac{1}{2}$ with each other and with the rotational motion splits each $|m_J|=1$ manifold into 8 states. The splittings are of order $\SI{10}{\mega\hertz}$, large enough that, to a good approximation, the tweezer light does not mix different hyperfine states. In this limit, the only degenerate pairs that split linearly with $I_{\rm tens}$ are those with $m_G=0$. There are two such pairs within each $|m_J|=1$ manifold, and they are suitable for our scheme.
In CaF, the near diagonal Franck-Condon factors to the $A^2\Pi$ and $B^2\Sigma$ states almost realise the ideal case of a single vibrational contribution to the polarizability from each electronic state. In the Appendix, we show that a suitable wavelength for the tensor tweezer is $\lambda_{\rm tens} \simeq \SI{545}{\nano\meter}$. Here, $\tilde{\alpha}^{(2)}\simeq\SI{6.3e-3}{\hertz\per\watt\meter\squared}$, $\alpha^{(0)}\simeq0$, and $R_{\rm ph} \simeq  \SI{0.1}{\per\second}$ per \si{\micro\kelvin} of trap depth.

\section{Summary} 

We have presented a general approach to building state-dependent tweezer traps for molecules based on their large tensor Stark shifts, and shown how they can be used to control the separation of molecules on a sub-wavelength scale. These ideas should work for a wide variety of molecules, and we have described how they can be implemented in practice for the $^{1}\Sigma$ and $^{2}\Sigma$ molecules already prepared in tweezers. The methods will be useful for studying collisions and reactions between molecules with control over all degrees of freedom including the wavefunction overlap. The reduced separation increases the dipole-dipole interaction between the molecules by at least a factor 100, bringing typical interaction strengths above 100~kHz. The enhancement can be even greater if an electric field is used to suppress inelastic collisions. Photon scattering rates in our scheme are not much larger than for separated tweezers and are unlikely to be a limitation. The enhanced dipolar interactions will transform the utility of polar molecules for quantum information processing and the exploration of many-body quantum systems.

\begin{acknowledgements}

We are grateful to Jeremy Hutson and Simon Cornish for helpful discussions. This work was supported by EPSRC grant EP/P01058X/1.

\end{acknowledgements}


\appendix*

\section{Details of polarizability and scattering rate calculations}
\label{sec:appendix}

We calculate the polarizabilities for light with angular frequency $\omega_{\rm L}$,
\begin{align}
    \alpha^{(0)} &= \frac{1}{3}(\alpha_\parallel + 2 \alpha_\perp),\\
    \alpha^{(2)} &= \frac{2}{3}(\alpha_\parallel - \alpha_\perp),
\end{align}
where 
\begin{align}
    \alpha_{\parallel(\perp)} &= \frac{1}{\hbar}\sum_j  \Bigg(\frac{\omega_{ij}-\omega_{\rm L}}{(\omega_{ij}-\omega_{\rm L})^2+\frac{\Gamma_j^2}{4}}\nonumber\\&\quad\quad+\frac{\omega_{ij}+\omega_{\rm L}}{(\omega_{ij}+\omega_{\rm L})^2+\frac{\Gamma_j^2}{4}}\Bigg)|\bra{i}d_{0(1)}\ket{j}|^2.
\end{align}
Here, $d_q$ is the $q$th component of the dipole moment operator in the frame of the molecule, $\Gamma_j$ is the decay rate from excited state $j$ and $\omega_{ij}$ is the transition angular frequency between $i$ and $j$. $i$ labels the lowest vibrational level of the ground electronic state, and $j$ stands for the quantum numbers $\{\eta',v'\}$ where $v'$ is the set of vibrational states within each electronic excited state $\eta'$. The matrix element $\bra{i}d_{q}\ket{j}$ is only non-zero for $\Sigma$ states when $q=0$, and for $\Pi$ states when $q=\pm 1$. It can be expressed as
\begin{equation}
    \bra{i}d_{q}\ket{j} = \bra{v=0}d_q^{X,\eta'}(R)\ket{v'}
    \label{eq:idqj}
\end{equation}
where $\vec{d}^{X,\eta'}(R)$ is the transition dipole moment as a function of the internuclear distance $R$, for the transition between X and $\eta'$.

Similarly, in the limit where the excited state fraction is small, the photon scattering rate is 
\begin{equation}
    R_{\rm ph} = \sum_j \frac{\Omega_{ij}^2}{4}\Bigg(\frac{1}{\omega_{ij}-\omega_{\rm L}}+\frac{1}{\omega_{ij}+\omega_{\rm L}}\Bigg)^2\Bigg(\frac{\omega_{\rm L}}{\omega_{ij}}\Bigg)^3\Gamma_j
\end{equation}
where $\Omega_{ij}$ is the Rabi frequency for the transition from ground state to excited state $j$ and this time the sum is over all excited states.

\begin{figure}
    \centering
    \includegraphics{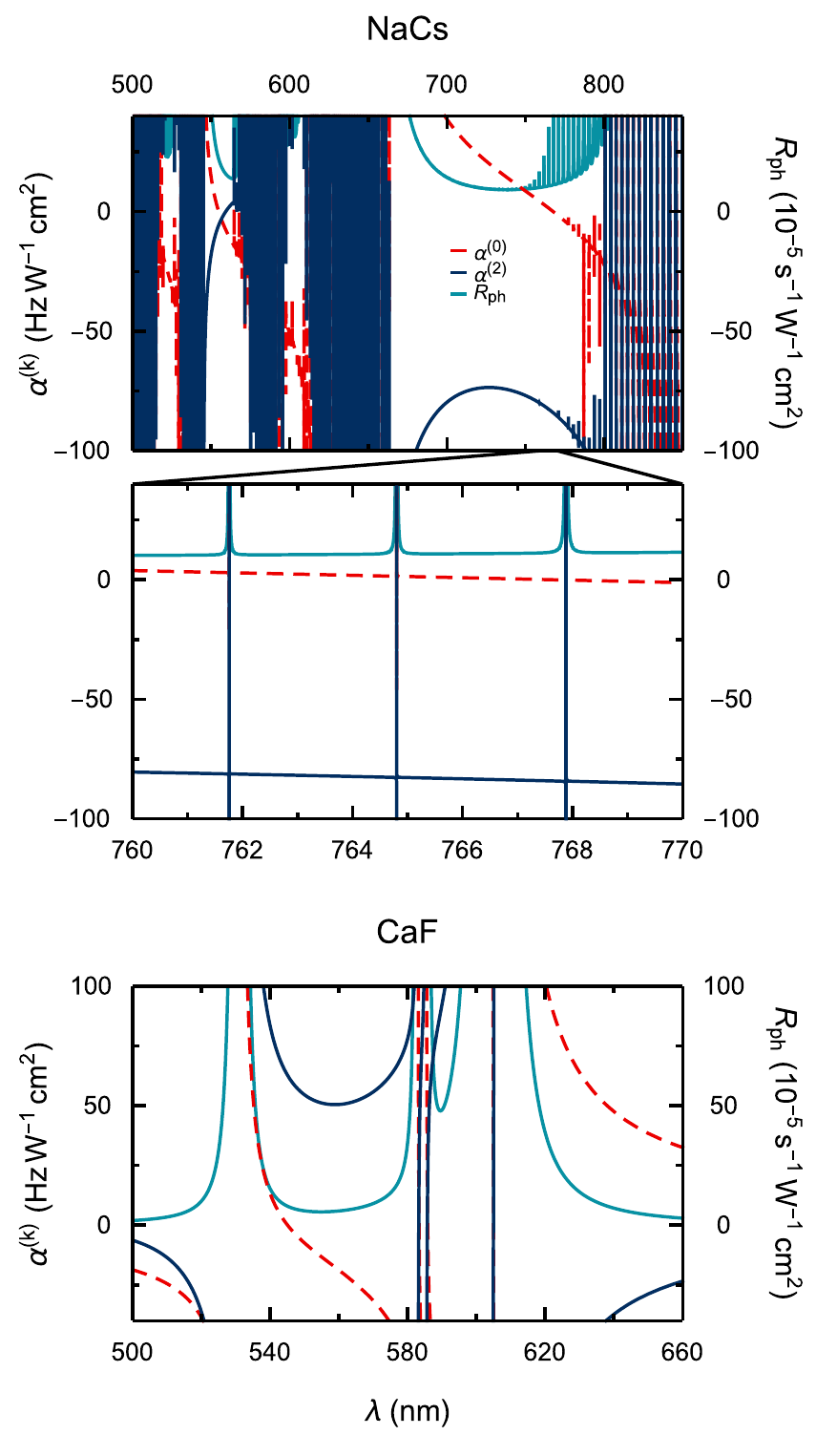}
    \caption{Calculated polarizabilities and scattering rates for $^{23}$Na$^{133}$Cs and $^{40}$Ca$^{19}$F.}
    \label{fig:pol-scat}
\end{figure}

\subsection{NaCs}
The upper part of Fig.~\ref{fig:pol-scat} shows our calculations of the polarizabilities and scattering rate in NaCs. In NaCs, many vibrational levels from each electronic excited state contribute to the polarizability and scattering rate. We calculate the set of energies and vibrational wavefunctions for each of the molecular potentials given in Ref.~\cite{Vexiau2017}. We then calculate the set of transition dipole moments using Eq.~(\ref{eq:idqj}) and $\vec{d}^{X,\eta'}(R)$ from Ref.~\cite{Vexiau2017}, and finally calculate the polarizability components and the scattering rate. We assume a lifetime of \SI{10}{\nano\second} for each of the excited states, following Ref.~\cite{Vexiau2017}. At \SI{767.64}{\nano\meter} we find $\tilde{\alpha}^{(0)}=\SI{2.3e-4}{\hertz\per\watt\centi\meter\squared}$, $\tilde{\alpha}^{(2)}=\SI{-84}{\hertz\per\watt\centi\meter\squared}$ and $R_{\rm ph}=\SI{1.2e-4}{\per\second\per\watt\centi\meter\squared}$. We note that although there are nearby transitions to highly excited vibrational levels of the $A^1\Sigma^+$ state, their wavefunction overlap with the ground state is so poor that it is possible to find wavelengths with very low scattering rate between them.

\subsection{CaF}
The lower part of Fig.~\ref{fig:pol-scat} shows our calculations of the polarizabilities and scattering rate in CaF. For CaF we assume that the polarizability and scattering rate are dominated by the $A^2\Pi$ state and the $B^2\Sigma^+$ state. We use the spectroscopic constants from Ref.~\cite{Kaledin1999} and the transition dipole moments from Refs.~\cite{Wall2008, Dagdigian1974}. The near-diagonal Franck-Condon factors from the ground state to each of these excited electronic states mean that it is only necessary to include the first three vibrational levels of each electronic state in the summation. At \SI{545.08}{\nano\meter} we find $\tilde{\alpha}^{(0)}=\SI{4.3e-5}{\hertz\per\watt\centi\meter\squared}$, $\tilde{\alpha}^{(2)}=\SI{63}{\hertz\per\watt\centi\meter\squared}$ and $R_{\rm ph}=\SI{7.9e-5}{\per\second\per\watt\centi\meter\squared}$.

\bibliography{references}

\end{document}